# RIMES: Embedding Interactive Multimedia Exercises in Lecture Videos

Juho Kim[1,2]   Elena L. Glassman[1,2]   Andrés Monroy-Hernández[1]   Meredith Ringel Morris[1]

[1]Microsoft Research, Redmond, WA, USA
[2]MIT CSAIL, Cambridge, MA, USA

{juhokim, elg}@mit.edu                    {andresmh, merrie}@microsoft.com

**ABSTRACT**
Teachers in conventional classrooms often ask learners to express themselves and show their thought processes by speaking out loud, drawing on a whiteboard, or even using physical objects. Despite the pedagogical value of such activities, interactive exercises available in most online learning platforms are constrained to multiple-choice and short answer questions. We introduce RIMES, a system for easily authoring, recording, and reviewing interactive multimedia exercises embedded in lecture videos. With RIMES, teachers can prompt learners to record their responses to an activity using video, audio, and inking while watching lecture videos. Teachers can then review and interact with all the learners' responses in an aggregated gallery. We evaluated RIMES with 19 teachers and 25 students. Teachers created a diverse set of activities across multiple subjects that tested deep conceptual and procedural knowledge. Teachers found the exercises useful for capturing students' thought processes, identifying misconceptions, and engaging students with content.

**Author Keywords**
Educational videos; online education; interactive exercises.

**ACM Classification Keywords**
H.5.1. Multimedia Information Systems: Video

**INTRODUCTION**
Lecture videos are gaining popularity in traditional classrooms, with many teachers using online lectures as supplementary materials, or even "flipping" their classes by assigning lecture videos to watch at home and spending class time for hands-on exercises and discussions [29]. Such blended learning models (e.g., flipped classrooms [29] and Small Private Online Courses (SPOCs) [13]) as well as online-only classes leverage videos specifically created for the course, accessible as open educational resources (OER) (e.g., MIT OpenCourseWare [ocw.mit.edu]), or available on large-scale online learning platforms that have garnered the attention of millions of learners recently [24,30] (e.g., Khan Academy, YouTube, and Massive Open Online Courses (MOOCs) [6]).

Most online video lecture systems, however, only provide disconnected and passive learning experiences. Lectures are commonly captured as videos, but interactive components that promote deeper understanding in classrooms – speaking out loud, drawing on a whiteboard, or using physical objects – are often lost. Furthermore, while videos and assessments are two of the primary features on online platforms [18], these components are often separate, potentially causing a cognitive overhead for students in linking information [22]. Many online computer programming courses include interactive coding exercises, but assessments in most other subject areas are constrained to multiple-choice and short answer questions that lack the interactivity and expressiveness common in classroom activities. Supporting richer modalities and activities for lecture videos requires significant technical expertise and resources that most teachers do not have.

Education research shows that higher interactivity in a video improves learning [35], and constructive and active learning (e.g., working out an answer in steps) outperforms passive learning (e.g., watching a video) [9]. Enabling multiple modalities such as drawing, moving, and speaking, in addition to typing, can provide learners with more opportunities for expressing ideas and learning deeply.

We introduce RIMES (Rich Interactive Multimedia Exercise System), a system for authoring, recording, and reviewing interactive multimedia exercises embedded in video lectures. The system's authoring interface allows teachers to create and embed these exercises in videos produced with Office Mix [officemix.com], a free plug-in for PowerPoint that turns slides into online videos with voice and webcam recordings of the presenter.

RIMES offers a new and expressive exercise widget for Office Mix, allowing for richer and more open-ended exercises than traditional quizzes. While watching the video, and without leaving their web browser, students are prompted to record their own responses, using their webcam, microphone, and inking (via mouse, touch, or stylus). Finally, RIMES aggregates all the students' responses in a gallery that lets teachers get an overview of all submissions as well as replay individual responses.

We evaluated the RIMES system and its workflow with 19 middle and high school teachers who authored video lectures with RIMES exercises, and 25 middle and high school students who recorded their responses to those same exercises. We also hired crowd workers on Amazon's Mechanical Turk to answer the exercises, simulating larger gallery sizes. Finally, the teachers returned to the lab to review the responses using the RIMES gallery.

Teachers created a wide variety of RIMES exercises across multiple subjects (math, sciences, English, and history), modalities (drawing, audio, and video), and question types (asking for open-ended responses, step-by-step explanations, and visual representations). Teachers reported that the system was effective at helping them identify misconceptions, capturing students' thought processes, and engaging students through formative assessment.

This paper makes the following contributions:

- A workflow for authoring, recording, and reviewing multimedia responses inside educational videos.

- A working prototype that implements the workflow as an extension to PowerPoint's Office Mix.

- A lab study with teachers and students showing how RIMES might be applied to the classroom.

## BACKGROUND AND RELATED WORK

Previous research on the value of interactive exercises and the technologies that support them was instrumental in the development of RIMES.

### Pedagogical Benefits of In-Video, Interactive Exercises

Constructivism and active learning theories emphasize the importance of the learner's active role in engaging with the learning material and the learning process: *"Students do more than just listen: They must read, write, discuss, or be engaged in solving problems"* [5]. Examples of active learning activities include group discussions, short written exercises, learning by teaching, and reacting to a video. Previous research has shown the pedagogical benefits of active learning over traditional lectures [14,27]. A meta-analysis of hundreds of research studies, dubbed ICAP [9], found that learning increases as engagement increases. For instance, simply watching a video is a *passive* activity, while answering in-video quizzes and open-ended exercises can promote *active* and *constructive* learning. ICAP ranked these activities as follows (from most to least learning impact): Interactive > Constructive > Active > Passive.

Another form of constructive learning is *self-explanation*, an activity that prompts learners to explain their answers and thinking [10]. It increases learning by encouraging students to make inferences and adjust mental models. A surprising research finding is that the learning benefits of self-explanations hold even without the presence of a teaching expert or any performance feedback [10]. Moreover, research has shown that these learning benefits extend to multimedia learning environments [34]. In terms of modality for explanation, students self-explained more frequently when asked to speak than when asked to type [16]. RIMES supports written and verbal self-explanations.

Previous research has also shown that interactive videos supporting random access and self-paced browsing improve learning [35] and task performance [17]. Tutored Videotape Instruction [15] reported improved learning when pausing videos every few minutes for discussion. RIMES supports such interactive activities for online videos. VITAL [26] allows students to clip and annotate on video segments, and further embed them in multimedia essays. While VITAL enables adding videos to an activity, RIMES takes an opposite approach by enabling adding activity to a video.

### Systems Supporting Interactive Online Learning

Many existing online learning platforms support in-video quizzes for flipped classrooms and MOOCs. Khan Academy [khanacademy.org] links supplementary videos to practice problems, and EDpuzzle [edpuzzle.com] and eduCanon [educanon.com] allow inserting quizzes into any video. MOOC platforms such as Udacity [udacity.com], Coursera [coursera.org], and edX [edx.org] commonly pause the video every few minutes and ask short questions to engage learners and test learning. Previous research has attempted to integrate video, comments, and assessments in a single view [22]. Short answer prompts for online learning have been shown to help learners self-correct misconceptions when asked to explain anomalies [33], and engage learners in tutorial videos while collaboratively summarizing them [31]. RIMES extends these systems by supporting rich media responses rather than multiple-choice or typed short answer questions. RIMES also captures the problem-solving process rather than just the outcome.

Some classroom technologies have used inking to support open-ended, freeform responses from students. Classroom Presenter [1] and Ubiquitous Presenter [32], for example, allow teachers and students to use digital ink on slides and share the annotated slides. Classroom Learning Partner [19] interprets handwritten answers and aggregates them into equivalent classes for built-in question types. These systems share many design goals with our work, but they are designed for use in traditional lecture settings; RIMES incorporates inking into asynchronous video learning, and allows teachers to author inking-oriented activities.

Recent systems in "learning at scale" settings attempt to support modalities beyond multiple-choice questions, such as open-ended short answers [3,7] and essay writing [21]. Improving social learning support is another thread of active research, via discussion forums, peer feedback and grading [25], chat rooms [11], and live video discussions [8]. RIMES contributes to this area of work with a new type of rich activity, namely recorded multimedia responses.

## DESIGN GOALS

This work aims to support more interactive and expressive exercises inside lecture videos. To discover the needs and challenges teachers have in using videos, we participated in a three-hour workshop at a local school with nine teachers (for grades 6–12) and three technical staff. They were asked to create lecture videos using the Office Mix plugin for PowerPoint, which turns slides into lesson videos with voice and video recordings per slide as well as exercise widgets such as multiple-choice or short answer questions. Observations and interviews with the teachers and the staff, as well as our review of prior research, led to the following high-level goals that informed the design of RIMES:

**Capture the thought process**. Multiple-choice or short answer questions are great for quickly aggregating responses, but they only capture the final outcome of the student's thought process. Teachers noted that simply checking against the right answer is not enough to ensure that students understand the material. Students might have answered correctly without really knowing why, or they might have missed one last trivial step when answering incorrectly. Teachers noted that in classrooms they walk around to listen to group discussions, and ask students to show their work in paper- or whiteboard-based exercises. However, for lecture videos that are not live, capturing the problem solving process is difficult.

**Enable multiple input modalities**. Many teachers found existing in-video exercises to be limiting, as they only allow students to type text or select from a pre-populated list. Some exercises naturally lend themselves to demonstrating, speaking, or drawing, most of which teachers use only inside classrooms. For flipped classrooms or purely online classes, supporting additional input modalities can enable more diverse and expressive exercises.

**Support customizable, open-ended exercises**. Many teachers mentioned that they would like to add interactive exercises to their videos or import external exercises, but they found the integration cost to be high. They often had specific requirements for tailoring exercises to their classes. Teachers expressed the need for adding their own commentary or exercises inside videos, which are not currently supported by most existing environments.

**Make it easy to integrate into existing practice**. While many teachers wanted to use videos in their classes, they pointed out having limited time to adopt new technologies. Only a few had experience editing and publishing videos, which suggests it is important for a video lecture system to work with what teachers are already familiar with. RIMES augments PowerPoint, a common lecture-delivery tool used by many teachers, to minimize integration issues.

## RIMES: RICH INTERACTIVE MULTIMEDIA EXERCISE SYSTEM FOR LECTURE VIDEOS

This section introduces RIMES, a system for teachers to author and manage multimedia exercises inside lecture videos, and for students to record their own responses while watching the videos. The following scenario illustrates a use case and a process for working with RIMES exercises.

### Usage Scenario

Allison is a middle school teacher looking to flip her English lecture on subject-verb matching. Rather than assigning a lecture video for her students to watch as homework, she wants to embed interactive exercises into her video. She expects to get students more excited about the material, and identify model answers and common misconceptions she can address during class time.

After deciding to add an interactive exercise after the slide with an example of a compound subject, Allison inserts a RIMES exercise into a blank slide; the slide now displays the RIMES authoring interface. She types in the instruction *("Record yourself as you identify the subject and the verb of each sentence. Circle the subject in green and the verb in red.")*, and uploads an image file of the sentences to show as the background. She records her lecture using the default features of Office Mix and publishes the video. She shares the web link to the video with her students.

Bobby, a student in Allison's class, opens the video link from a tablet computer at his house. As he watches the video, the video pauses after the compound subject example, and asks him to identify the subject and verb from two example sentences by drawing on them and voice-recording his answer. After reflecting for a while, he clicks "Record." He uses his finger to circle the subject in green and the verb in red, and verbally describes the rationale behind his selections. He clicks on "Stop" to submit his response. The video then advances to the next slide.

Allison opens the RIMES gallery interface before class to review all students' responses. By visually scanning thumbnails of all the responses (sentences marked up with green and red ink), she is happy with the overall quality of the answers. But she notices a few incorrect answers, and clicks on one submitted by Ryan to listen to his response. The player plays back Ryan's voice and pen strokes. Allison discovers that Ryan did not correctly understand the compound subject because he said, *"a subject should be a noun…"* Browsing the gallery, she finds that two other students made the same mistake and notes to remind students of this common error in the next class.

### Authoring

The RIMES authoring interface (Figure 1) lets the teacher embed a custom RIMES exercise into any slide. Inserting a RIMES object in PowerPoint renders the authoring interface directly into the current slide. It contains the following options for customizing the exercise:

- **Instructions**: the text prompt that students see when the video pauses and the RIMES exercise pops up.

- **Time limit**: the expected maximum length of a student response, useful for students in planning their response and for bounding media storage needs.

- **Input mode**: a student response can use any combination of video, audio, and drawing. Available

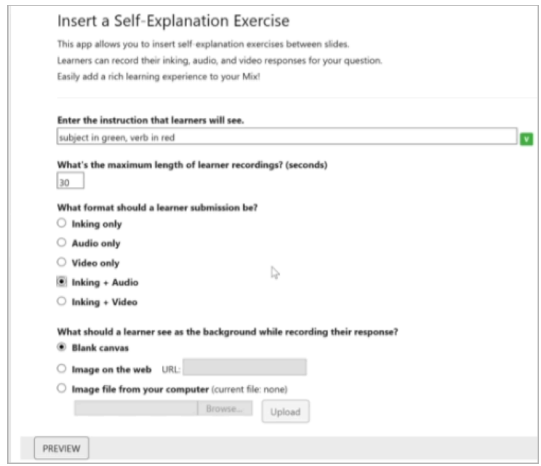

**Figure 1. The RIMES authoring interface allows the teacher to create and customize RIMES exercises by directly editing a PowerPoint slide.**

modes are: "Inking only," "Audio only," "Video only," "Inking + Audio," and "Inking + Video."

- **Background image**: an optional background image that students can reference or mark on. The teacher can add a link to any web image, or upload their own.

The teacher can click on a "Preview" button within the RIMES slide to check what the recording view will look like for students. The teacher can move between the preview and edit modes to update the exercise. Once a video including a RIMES exercise is published, the RIMES exercise is part of the video playback sequence. The video automatically pauses when the play control reaches the slide with the RIMES exercise. The Office Mix tool already supports the video publication and playback environment.

**Recording**

While watching a video, students are prompted to record their own response using video, audio, and/or drawing when they encounter a RIMES slide (Figure 2). They can explain their solution using multiple modalities inside a web browser without having to install any plugins.

The recording interface starts capturing any of the enabled modalities (video, audio, drawing) once the student clicks the "Record" button. When inking is enabled, the sidebar shows a palette for changing the pen size and color. The student can use touch or a stylus if available, or a mouse if not. When video is enabled, the student sees a visual preview in real-time. When the student is done recording, she clicks "Stop." She is given options to either re-record or submit. Once she submits, the submission dialog box opens, asking her to rate how confident she is about her response and how helpful the RIMES exercise was. The teacher can refer to these scores when reviewing students' responses, as research suggests that self-efficacy and confidence are effective indicators of learning [2].

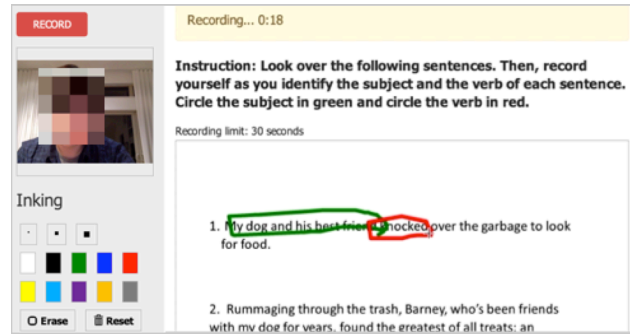

**Figure 2. The RIMES recording interface in a web browser. The student's face is blurred from the screenshot for privacy.**

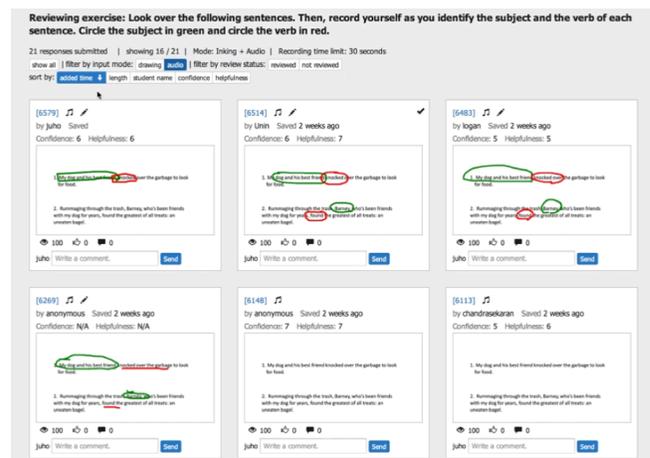

**Figure 3. The RIMES reviewing interface shows a gallery of all responses to a RIMES exercise. Various sorting and filtering options help the teacher with the reviewing process.**

Once a recording is submitted, the RIMES backend post-processes each recording for improved navigation in the gallery. It detects responses with silent audio or no pen input, and labels them as "no audio" or "no inking," respectively.

**Reviewing**

The gallery for each RIMES exercise (Figure 3) is automatically populated as the students submit their responses. The gallery renders each response in a card-style layout, and displays thumbnails of the final drawing canvas or video recording if available, student name, helpfulness and confidence ratings, recorded length, and captured input modes. The gallery is designed to support the following use cases: 1) get a quick overview of all student responses, 2) easily identify model answers and misconceptions, and 3) replay a particular student's response to better understand his thought process. The specific gallery features are:

- **Sorting**: the teacher can sort all responses using various sorting options, such as submission timestamp, recording length, student name, and students' self-reported confidence and helpfulness ratings. These options help teachers prioritize their reviewing order.

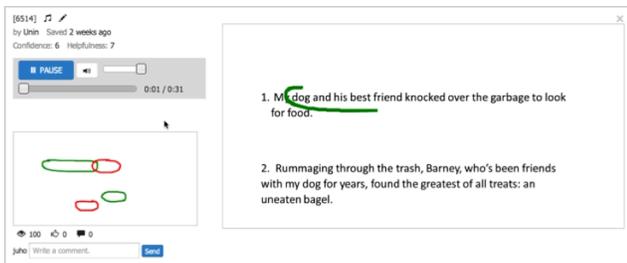

**Figure 4. Clicking on a response in the gallery opens the player, letting the teacher synchronously replay the captured video, audio, or inking stroke sequences.**

- **Filtering**: the teacher can selectively view responses that match certain filtering criteria. Available options include input mode (e.g., selecting "drawing" displays all responses that include pen strokes), and review status (e.g., selecting "not reviewed" displays all responses that the teacher has not yet played).
- **Playback** (Figure 4): the teacher can replay the process in which a response was recorded. Based on the input modes captured, the player provides a synchronized playback of all modalities used in a response.
- **Social interactions**: the teacher can either "like" or leave comments for each response, both as feedback to students and as a note to herself. Students given access to the gallery can read comments and reply as well.

**Implementation**

We implemented RIMES using standard web technologies: HTML5, CSS3, and JavaScript. For audio and video recording, the WebRTC protocol is used. To reduce the network bandwidth required in recording, canvas strokes are not captured as video but as JSON with per-stroke timestamp, action type, and action coordinates. The canvas playback reads the JSON and replays the strokes in animation, using the *requestAnimationFrame* method.

We implemented our conceptual workflow for managing multimedia in-video exercises in the context of PowerPoint and Office Mix. The general idea can be implemented for other video platforms (e.g., YouTube) with simple player customizations for RIMES. The authoring interface is a Microsoft Office application, which uses the *labs.js* [labsjs.blob.core.windows.net/sdk/index.html] library. This library allows a standalone web application to be recognized by PowerPoint. This integration allows the RIMES authoring interface to open inside PowerPoint, and the recording interface to open RIMES exercises inside the Office Mix player on its portal [officemix.com].

## EVALUATION: TESTING THE RIMES WORKFLOW

We conducted a three-part user study to evaluate the three-stage RIMES workflow end-to-end. We designed our study to allow us to address the following questions:

| Teaching Subject | Count |
|---|---|
| Math | 7 |
| English | 5 |
| Science | 4 |
| Social Science | 3 |
| Technology & Computers | 4 |

| Teaching Grade | 6 | 7 | 8 | 9 | 10 | 11 | 12 |
|---|---|---|---|---|---|---|---|
| Count | 4 | 4 | 4 | 9 | 8 | 8 | 8 |

**Table 1. Subjects and grades taught by the 19 teachers in Study 1. Multiple responses were allowed.**

- **Authoring**: can teachers create a variety of RIMES exercises spanning multiple subjects, input modes, and question types?
- **Recording**: can students effectively record their responses to a teacher's RIMES exercise?
- **Reviewing**: can teachers browse students' responses to their RIMES exercises to get an overview of responses, identify model answers and misconceptions, and understand students' thought processes?

## STUDY 1. TEACHERS AUTHOR RIMES EXERCISES

The first study focused on the RIMES creation experience.

**Participants**

We recruited 19 middle and high school teachers (13 female, 5 male, 1 preferred not to specify, mean age=49, σ=11, max=68, min=27, T1-T19) from a large, U.S. metropolitan area. They had 21 years of teaching experience on average, with 14 of them teaching at a public school and 5 at a private school at the time the study was conducted. Subjects and grades they teach spanned a wide range (Table 1). Ten teachers indicated that they had made instructional videos before, using software such as Camtasia, Adobe Premiere, and iMovie. Seven indicated that they had flipped their classrooms before. In a 5-point scale question (1: not familiar, 5: very familiar), most teachers said they were familiar with PowerPoint (mean=3.8, σ=0.6). Five had prior experience with Office Mix. Participants were offered a gratuity for their time.

**Procedure**

Each study session was a three-hour workshop in which teachers were asked to create one or two educational videos using RIMES and Office Mix inside PowerPoint. We asked them to bring in their own existing slides as initial materials to create a realistic lecture video. Teachers were provided with a Windows tablet, webcam, external mouse, and stylus pen. An external microphone was available upon request for improved audio quality.

The session started with a pre-questionnaire asking participants about demographic information as well as teaching and technical experiences. After a tutorial on using

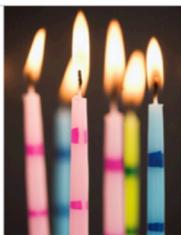
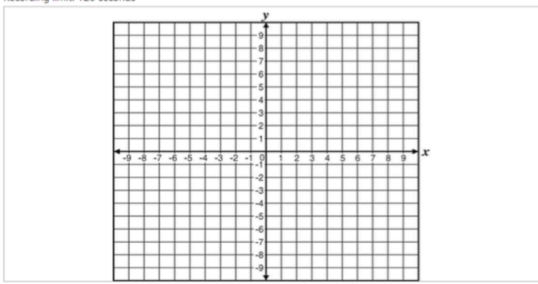

**Figure 5. Example RIMES exercises teachers created in Study 1: a) explain chemical properties of a candle while circling relevant parts, b) use color marking to identify imagery patterns in a poem, c) draw a graph on a chart grid.**

Office Mix and authoring RIMES exercises, teachers spent two hours creating videos. We asked them to make short videos (five minutes or shorter) and to include at least one RIMES exercise in each video they created. Researchers provided technical support when requested. After publishing each video to the Office Mix portal, teachers were asked to indicate the self-rated quality and target grade level of their video. The session concluded with a questionnaire about the pedagogical implications and usability of the RIMES tool.

After all sessions were complete, researchers manually selected one video and one RIMES exercise in the video from each teacher to assign to students in the second study. For a deeper understanding of the RIMES exercises created, researchers coded all RIMES exercises by subject area, as well as by the knowledge type and the level of cognitive process they test, using the revised version of Bloom's Taxonomy [20]. The original Bloom's Taxonomy [4] and the revised version [20] are commonly used frameworks to evaluate the level of learning assessed in exams and tests [36]. Two researchers independently coded each RIMES exercise by the knowledge and cognitive process dimensions. When a researcher assigned multiple codes to an exercise, the highest level was selected as the final label. Inter-rater reliability was high for both the knowledge (Cohen's κ [12]=0.83) and cognitive process (Cohen's κ=0.72) dimensions. For final dimensional labels (Figure 6), we took a higher-level value of the two researchers' labels when they did not match.

**Results**

The 19 teachers created 36 videos and 69 RIMES exercises (mean=1.9 videos per teacher, σ=1.4). The videos were, on average, 3.6 minutes long (σ=2.4), and had 10.4 slides (σ=5.3). Teachers used a background image in 41% of the RIMES exercises (28/69), such as a chart area for math, text to analyze for English, and a photo for science (Figure 5).

The RIMES exercises spanned multiple subjects, input modes, and knowledge and cognitive processes. Subject areas in the exercises included math (e.g., "Use substitution to solve the following system."), science (e.g., "Draw a picture of an organism showing one or more of the characteristics of life."), English (e.g., "When do we use capitalization?"), and history (e.g., "Explain why murdering a king was considered a disruption of order in the Renaissance world.").

Teachers added inking to a vast majority of RIMES exercises (60/69), while audio (37/69) and video (12/69) were used less frequently. The most common input mode was inking combined with audio (33/69), which gave students an opportunity to verbally explain their answer while drawing on a canvas or background image.

A qualitative analysis of the exercises using the revised Bloom's taxonomy reveals that the RIMES exercises require knowledge and cognitive processes beyond factual knowledge and rote memorization. In the knowledge dimension (Figure 6, top), only 23% of the exercises focused on factual knowledge, while 43% tested conceptual knowledge and 30% procedural knowledge. In the cognitive process dimension (Figure 6, bottom), *Remember* (e.g., "What three keys do we use to take a screenshot?"), the lowest level process, comprised 14%, while many exercises addressed higher-level processes such as *Understand* (30%, e.g., "Why did the regular coke sink and the diet coke float? What ideas do you have to explain this phenomenon?"), *Apply* (25%, e.g., "In the space below, please show your work that will allow you to find the weight of one math book."), and *Create* (17%, e.g., "Sketch a picture that represents how a Renaiassance [sic] person would see the relationship between a star, a flower and a mole."). In the ICAP model [9], RIMES exercises mostly fall under the Constructive and Active categories beyond Passive.

Teachers' responses to RIMES were mostly positive. On a 7-point Likert scale (1: strongly disagree, 7: strongly agree),

teachers found the RIMES recording interface to be easy (6.3) and enjoyable to use (6.4). They thought they were able to effectively add RIMES exercises (6.5), and most were willing to use it in the future (6.7). Many thought the rich media support could lead to a better understanding of students' thought processes: *"I really like that students can draw or respond to my prompts orally. I think that will really give me a window into their thinking."* [T19]

All but two teachers agreed that RIMES exercises would be pedagogically valuable. Teachers compared RIMES exercises against multiple-choice or short answer questions: *"Given multiple choice, many students just guess instead of working out a solution. Many of my students don't feel comfortable answering short answer options in a Math class. RIMES just works better."* [T3]

**STUDY 2. STUDENTS RECORD RESPONSES**

The second study was designed to test the recording interface, see students' reactions to RIMES, and collect responses for the teachers to review in the next phase.

**Participants**

We invited 25 middle and high school students from a large, U.S. metropolitan area to our lab (15 female, 9 male, 1 preferred not to specify, mean age=14.7, $\sigma$=1.4, max=17, min=13, S1-S25). Nine indicated that their teachers occasionally assign videos to watch at home. Participants were offered a gratuity for their time.

**Procedure**

In a one-hour session, students first answered a pre-questionnaire that obtained demographic information and feedback on students' video learning experiences. Researchers then gave a tutorial on the Office Mix video player and the RIMES recording interface, giving students a chance to practice recording. Students were provided with a Windows tablet, webcam, external mouse, and stylus pen.

Students then watched two videos created by teachers in Study 1. We ensured that students' grade level matched the target audience level of the exercise, which the teachers specified a priori. While watching each video, students recorded their responses to RIMES exercises as they encountered them. Depending on the remaining time, 21 out of 25 students had a chance to look at the gallery interface and shared their thoughts on social interactions around student responses. The session concluded with a post-questionnaire on the usability of the recording interface and students' reaction to sharing their responses with other students and seeing others' responses.

**Supplemental Turker Responses**

We additionally recruited crowd workers on Mechanical Turk to record responses to the RIMES exercises. Although most Turkers were adults and beyond the target age range for most of the RIMES exercises, we used Mechanical Turk for two reasons. First, we wanted to test the usability of the recording interface with remote users who have diverse computing environments (Turkers used their own computers and input devices for recording responses). Second, we wanted to populate the gallery for each RIMES exercise with the number of responses comparable to that of a real classroom (20–30), which the 25 in-lab students watching two videos each could not fully populate. We created a total of 574 HITs (30.2 per RIMES exercise), with an average hourly wage of $8.98 over the course of 10 days. HIT instructions told Turkers they would need to watch a provided educational video and complete any exercises within the video. The HIT also included a few brief demographic questions, which showed that 46% of the Turkers were 24 or younger, 62% had a college degree or higher level of education, and 47% were female.

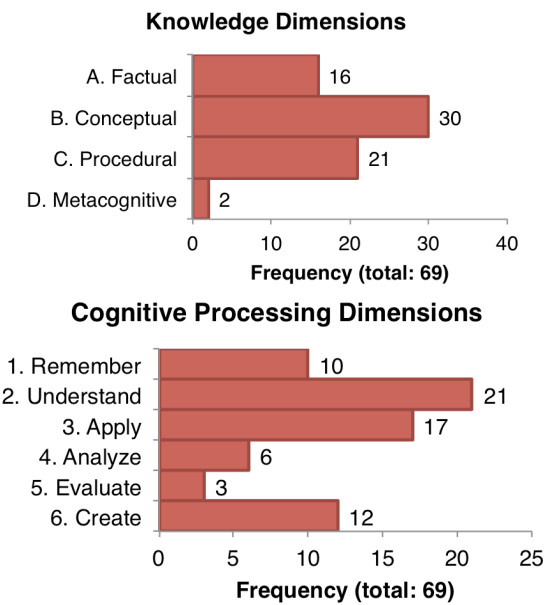

Figure 6. Frequency of RIMES exercises' knowledge (top) and cognitive process (bottom) dimensions using the revised Bloom's Taxonomy [20].

**Results**

Students' comments about the recording experience were generally positive. When questioned using a 7-point Likert scale (1: strongly disagree, 7: strongly agree), students found the recording interface to be easy (5.5) and somewhat enjoyable (4.8) to use. S1 commented, *"I liked being able to interact in the lesson, rather than just taking notes or listening."* They thought they could effectively add their responses (5.9), found the exercises to be engaging (5.5), and were willing to use the interface in the future (5.4). Many found the high interactivity of RIMES exercises helpful: *"It made sure you could actually learn it and if you were paying attention or not."* [S10] On the other hand, several students found self-recording to be awkward, the time limit to be too short (e.g., *"the time limits made it sometimes hard to complete your full thought and drawing"* [S14]), and the canvas size to be too small.

When asked about sharing responses with others in 4-point scale questions (1: not at all, 4: highly interested), students expressed interest in seeing other responses (3.0), nothing that they saw value in being able to see *"different ways of solving the same problem"* [S19], and comparing their answer against others: *"If some had similar ideas to mine, or what they did better than me so I could learn from them."* [S3] They were also moderately comfortable with other students seeing their responses (2.7). Primary reasons for a willingness to share their responses included helping others better understand the lesson and returning the help they got from seeing others' responses. On the other hand, some wanted to ensure anonymity and privacy, which future designs should carefully address.

## STUDY 3. TEACHERS REVIEW RESPONSES

The final study tested the gallery interface, where teachers reviewed the responses to their RIMES exercises.

### Procedure

The same 19 teachers in Study 1 returned to the lab after about 10 days to review collected responses to their RIMES exercises. They used the RIMES gallery interface, populated with 20 to 30 responses to one of their RIMES exercises submitted by both the Study 2 students and Turkers. After a tutorial on the interface, teachers freely explored the gallery, spending 10 to 20 minutes reviewing the responses while thinking aloud. A post-questionnaire asked teachers about the usability of the gallery interface, as well as how it might support their teaching practices.

### Results

We asked teachers how often they would be willing to create videos and use RIMES in them. All 19 teachers indicated they would be willing to use RIMES: 10 for every video, 6 no more than once a week, and 3 no more than once a month. Of the 10 who answered every video, 2 said they want to make a video every day, 6 every week, and 2 every month. While lab study responses may introduce the Hawthorne effect, teachers acknowledged the need for and value of interactive exercises, and many asked when RIMES would be publically available.

On 5-point Likert scale questions (1: strongly disagree, 5: strongly agree) about the gallery experience, teachers rated the gallery interface as easy to use (4.8), useful (4.5), and interesting (4.3). We organize additional findings by the three key goals the gallery supported:

#### Get an overview

Teachers thought the gallery made it easy to get an overview of student responses (4.3). T4 described, *"I like that all responses are on a single page and viewable without having to click anything."*

#### Identify good answers and misconceptions

Using the gallery, teachers answered that they were able to identify good answers (4.4) and misconceptions (4.2), and that it was easy to compare and contrast multiple responses (4.1). Because many teachers mentioned that finding time to review many lengthy responses could be a practical barrier to applying RIMES in their classrooms, it is encouraging that most teachers felt the gallery interface helped them quickly identify noticeable ones. T9 said, *"I liked how quickly [sic] it was to get a sense of student understanding and misunderstanding."* Many teachers applied sorting by helpfulness and confidence and paid special attention to those with low self-ratings: *"Sorting by helpfulness and confidence let me immediately identify people who thought they were struggling."* [T8]

#### Understand the thought process

Teachers replied that the gallery helped them understand students' thought processes (4.2). A common review strategy was to first skim all the responses to get an overview and pick a few noteworthy ones (normally outliers, such as incorrect, long, or visually different answers) for a detailed playback: *"The ability to see the time-lapse of the drawing helped me understand the thinking behind the drawings."* [T12]

Some teachers identified *why* the student reached the final answer. In a math exercise on functions that asked students to find outputs to various input values (shown below),

$$f(x) = x + 4$$
$$f(3) = ?\quad f(2) = ?\quad f(-10) = ?$$

the teacher played back a response with incorrect answers (6, 4, and -20 instead of 7, 6, and -6, shown below).

$$2\ f(x) = x + 4\quad 8$$
$$f(3) = 6\quad f(2) = ?\ 4\quad f(-10) = ?\ -20$$

While listening to the student's explanation, T18 said, *"Aha! This student thought this function doubles the input based on his test with f(4) that gave him 8. That's why he doubled all inputs. This is a common mistake students make when they first learn about functions."* In the gallery, there were a couple other responses that had this set of incorrect answers (shown below).

$$2\ f(x) = x + 4$$
$$f(3) = 6\ f(2) = ?\ 4\ f(-10) = ?\ -20$$

$$f(x) = x + 4$$
$$f(3) = ?\quad f(2) = ?\quad f(-10) = ?$$
$$f(3) = \quad ?\quad\quad F(-10)$$
$$3 \times 2\quad 2 \times 2\quad \times 2$$

T5 described the benefit of the playback, *"I believe that by listening to the responses of each student we truly get a realistic view of student understanding of concepts presented. It will definitely help teachers to know which students need assistance in understanding the concept."*

## DISCUSSION AND FUTURE WORK

We discuss several practical issues surrounding RIMES, including the choice of input, real-life applications, and the limitations of our findings.

**Input Modes**
The multiple input modes in RIMES serve complementary purposes, thus expanding the spectrum of exercises possible with RIMES. Drawing (inking) helped students visualize their thinking, and enabled richer, more expressive descriptions not possible with typing: *"[Drawing] made obvious what students imagined in their heads about the material and the ideas."* [T12] Audio-recording open-ended explanations *"forces the kids to be clear and articulate,"* [T18] and works better for certain students: *"There are always kids who do better [than writing] while explaining w/ audio."* [T8] Several teachers who required video recording appreciated that students' facial expression and body language were captured. A student in a physics exercise instructed to calculate kinetic energy showed his calculator screen in the video, to which T7 reacted, *"This is very nice as a teacher because this is where many errors are made."* Video, however, did not work for everyone. Some students expressed concern about anonymity and recording anxiety. Both for audio and video, several teachers indicated the difficulty of reviewing them, because they are more tedious to review than static answers that lend themselves to easy skimming.

**Possible Applications of RIMES**
When asked about the potential practical applications of RIMES, teachers suggested that RIMES would be beneficial for flipped classrooms, SPOCs [13], remedial lectures, or make-up lectures for students who missed a day. RIMES can also be useful in identifying and helping struggling students: *"I would also have a follow-up set of slides for remediation if needed for those students still having difficulty."* [T7] Distance learning and online-only classes can adopt RIMES, with frequent in-video exercises and recording opportunities increasing student engagement.

Rather than replacing existing summative assessment (e.g., tests) and easy-to-grade quizzes (e.g., multiple-choice questions), teachers indicated they would use RIMES as a formative assessment tool, through which they could refine their lectures and provide qualitative feedback to students [28]. Many teachers mentioned that they would not want to assign grades to RIMES responses for this reason.

**Challenges in Adopting RIMES**
We share challenges and design lessons for future designers and practitioners. Firstly, students' experience needs to be carefully designed because of recording anxiety, different preferences in input modalities, and preferences for socially sharing their response. Supporting re-recording, preview, and multiple input modes in RIMES mitigated some of these challenges. Also, teachers felt that RIMES exercises are more suitable for formative assessment, indicating that RIMES should be used in conjunction with other assessment tools. Giving guidance on how teachers can effectively use RIMES is another challenge; creating online galleries where teachers can share successful activities and re-use others' exercises will be important to adoption.

**Limitations**
Our evaluation used a mixed-methods approach, combining many data sources: questionnaires, observations, think-alouds, interviews, and qualitative analysis of RIMES artifacts. While this subjective data helped us answer how teachers and students would use and react to RIMES, it lacks conclusive evidence that RIMES leads to meaningful pedagogical improvements. Future work will incorporate quantitative measures such as click logs or learning gains.

A limitation of the study is that we recruited paid adult crowd workers to simulate online learners. While teachers confirmed that the responses were not noticeably different from what they normally see in their classes, this setup might have missed response patterns unique to the original target population. Also, due to practical recruiting constraints, the age-appropriate students we recruited for the lab portion of the study were not necessarily in the regular class of the teacher whose lecture they viewed. The dynamics of response recording and reviewing may vary when teachers and students interact on a regular basis. Since continuous feedback plays an important role in learning [23], and many teachers left comments for students using the gallery, a follow-up study focusing on feedback on RIMES would be valuable. We aim to publicly release RIMES, which may help us understand engagement and interaction dynamics not observed in the lab study.

**FUTURE WORK**
We plan to investigate the impact of using RIMES in classrooms in a longer-term, live deployment. Research questions include: How do the lab study findings translate to real teacher practices? How do the different interactivity and exercise types affect students' learning? How to help teachers save time in reviewing and providing feedback?

Future work will explore ways for students to interact with other students' responses to increase potential peer learning benefits. We also plan to support the collaborative RIMES experience in which multiple students record their discussion or outcomes achieved from teamwork.

**CONCLUSION**
We introduced RIMES, a system that allows teachers to easily create interactive multimedia exercises embedded within online lecture videos. Students can record audio, video, and ink-based answers from their web browsers, and teachers can efficiently review the responses. A three-part evaluation with $6^{th} - 12^{th}$ grade teachers and students found that the RIMES interface and workflow were usable and compelling. The teachers created exercises for a diverse array of subjects, involving diverse knowledge and cognitive processing dimensions. These studies illustrated how RIMES could facilitate creating online lectures that move beyond passive learning and simple assessments. RIMES presents a model for designing usable tools that empower teachers and students to create engaging and interactive content for online learning.

**ACKNOWLEDGMENTS**

The authors would like to thank Anoop Gupta, Kurt Berglund, Sasa Junuzovic, and the Office Mix team for their feedback and support.

**REFERENCES**
1. Anderson, R., Anderson, R., Davis, P., et al. Classroom presenter: Enhancing interactive education with digital ink. *Computer 40*, 9 (2007), 56–61.
2. Bandura, A. Self-efficacy: toward a unifying theory of behavioral change. *Psychological review 84*, 2 (1977).
3. Basu, S., Jacobs, C., and Vanderwende, L. Powergrading: a Clustering Approach to Amplify Human Effort for Short Answer Grading. *TACL 1*, (2013), 391–402.
4. Bloom, B. S. and Krathwohl, D.R. Taxonomy of educational objectives: The classification of educational goals. Handbook I: Cognitive domain. (1956).
5. Bonwell, C. C. and Eison, J.A. *Active Learning: Creating Excitement in the Classroom. 1991 ASHE-ERIC Higher Education Reports.* ERIC, 1991.
6. Breslow, L., Pritchard, D.E., DeBoer, J., Stump, G.S., Ho, A.D., and Seaton, D.T. Studying learning in the worldwide classroom: Research into edX's first MOOC. *Research & Practice in Assessment 8*, (2013), 13–25.
7. Brooks, M., Basu, S., Jacobs, C., and Vanderwende, L. Divide and correct: using clusters to grade short answers at scale. *Learning at Scale '14*, ACM (2014), 89–98.
8. Cambre, J., Kulkarni, C., Bernstein, M.S., and Klemmer, S.R. Talkabout: small-group discussions in massive global classes. *Learning at Scale '14*, ACM (2014).
9. Chi, M.T. Active-constructive-interactive: A conceptual framework for differentiating learning activities. *Topics in Cognitive Science 1*, 1 (2009), 73–105.
10. Chi, M.T., Bassok, M., Lewis, M.W., Reimann, P., and Glaser, R. Self-explanations: How students study and use examples in learning to solve problems. *Cognitive science 13*, 2 (1989), 145–182.
11. Coetzee, D., Fox, A., Hearst, M.A., and Hartmann, B. Chatrooms in MOOCs: All Talk and No Action. *Learning at Scale '14*, ACM (2014), 127–136.
12. Cohen, J. A coefficient of agreement for nominal scales. *Educational and Psychological Measurement 20*, (1960), 37–46.
13. Fox, A. From MOOCs to SPOCs. *Communications of the ACM 56*, 12 (2013), 38–40.
14. Freeman, S., Eddy, S.L., McDonough, M., et al. Active learning increases student performance in science, engineering, and mathematics. *PNAS 111*, 23 (2014).
15. Gibbons, J.F. Tutored Videotape Instruction. (1977).
16. Hausmann, R.G. and Chi, M.H. Can a computer interface support self-explaining. *Cognitive Technology 7*, 1 (2002), 4–14.
17. Kim, J., Nguyen, P.T., Weir, S., Guo, P.J., Miller, R.C., and Gajos, K.Z. Crowdsourcing Step-by-step Information Extraction to Enhance Existing How-to Videos. *CHI'14*, ACM (2014), 4017–4026.
18. Kizilcec, R.F., Piech, C., and Schneider, E. Deconstructing Disengagement: Analyzing Learner Subpopulations in Massive Open Online Courses. *LAK'13*, ACM (2013), 170–179.
19. Koile, K., Chevalier, K., Rbeiz, M., et al. Supporting Feedback and Assessment of Digital Ink Answers to In-class Exercises. *IAAI'07*, (2007), 1787–1794.
20. Krathwohl, D.R. A revision of Bloom's taxonomy: An overview. *Theory into practice 41*, 4 (2002), 212–218.
21. Markoff, J. Essay-grading software offers professors a break. *The New York Times Apr. 4, 2013*, (2013).
22. Monserrat, T.-J.K.P., Li, Y., Zhao, S., and Cao, X. L.IVE: An Integrated Interactive Video-based Learning Environment. *CHI'14*, ACM (2014), 3399–3402.
23. Mory, E.H. Feedback research revisited. *Handbook of research on educational communications and technology 2*, (2004), 745–783.
24. Noer, M. One Man, One Computer, 10 Million Students: How Khan Academy Is Reinventing Education. *Forbes*, Nov. 19, 2012 (2012).
25. Piech, C., Huang, J., Chen, Z., Do, C., Ng, A., and Koller, D. Tuned models of peer assessment in MOOCs. *arXiv preprint arXiv:1307.2579*, (2013).
26. Preston, M., Campbell, G., Ginsburg, H., et al. Developing New Tools for Video Analysis and Communication to Promote Critical Thinking. *EdMedia'05*, (2005), 4357–4364.
27. Prince, M. Does Active Learning Work? A Review of the Research. *J. of Engineering Education 93*, 3 (2004).
28. Sadler, D.R. Formative assessment and the design of instructional systems. *Instructional science 18*, 2 (1989).
29. Sams, A. and Bergmann, J. Flip your classroom: Reach every student in every class every day. *International Society for Technology in Education (ISTE)*, (2012).
30. Shumski, D. MOOCs by the numbers: How do EdX, Coursera and Udacity stack up? *Education Dive August 15, 2013*, (2013).
31. Weir, S., Kim, J., Gajos, K.Z., and Miller, R.C. Learnersourcing Subgoal Labels for How-to Videos. *CSCW'15*, ACM (2015).
32. Wilkerson, M., Griswold, W.G., and Simon, B. Ubiquitous presenter: increasing student access and control in a digital lecturing environment. *ACM SIGCSE Bulletin 37*, 1 (2005), 116–120.
33. Williams, J.J., Kovacs, G., Walker, C., Maldonado, S., and Lombrozo, T. Learning Online via Prompts to Explain. *CHI EA'14*, ACM (2014), 2269–2274.
34. Wylie, R. and Chi, M.T. 17 The Self-Explanation Principle in Multimedia Learning. *The Cambridge Handbook of Multimedia Learning*, (2014), 413.
35. Zhang, D., Zhou, L., Briggs, R.O., and Nunamaker Jr, J.F. Instructional video in e-learning: Assessing the impact of interactive video on learning effectiveness. *Information & Management 43*, 1 (2006), 15–27.
36. Zheng, A.Y., Lawhorn, J.K., Lumley, T., and Freeman, S. ASSESSMENT: Application of Bloom's Taxonomy Debunks the "MCAT Myth." *Science 319*, 5862 (2008).